\documentclass[figures,preprint]{epl}   % Format fuer Europhys.Lett. 
\usepackage{amsfonts}
\usepackage{amssymb}
\usepackage{amsbsy}
\usepackage{graphics}
\usepackage{psfig}

\input epsf.sty

\newcommand{\BEQ}{\begin{equation}}     % Gleichungen Anfang ..
\newcommand{\BEA}{\begin{eqnarray}}
\newcommand{\EEQ}{\end{equation}}       % .. und Ende
\newcommand{\EEA}{\end{eqnarray}}
\newcommand{\eps}{\varepsilon}          % epsilon
\newcommand{\D}{{\rm d}}                % gerades d fuer Ableitungen
\renewcommand{\vec}[1]{\boldsymbol{#1}} % Vektoren fettgedruckt

\title{Ageing and dynamical scaling in the critical Ising spin glass}
\shorttitle{Ageing in the critical Ising spin glass}

\author{Malte Henkel\inst{1} and Michel Pleimling\inst{2}}

\institute{
\inst{1}Laboratoire de Physique des Mat\'eriaux (CNRS UMR 7556), 
Universit\'e Henri Poincar\'e \\ Nancy I, B.P. 239,
F -- 54506 Vand{\oe}uvre l\`es Nancy Cedex, France\\
\inst{2} Institut f\"ur Theoretische Physik I, 
Universit\"at Erlangen-N\"urnberg,\\
D -- 91058 Erlangen, Germany}
\pacs{05.70.Ln}{Nonequilibrium and irreversible thermodynamics}
\pacs{75.50.Lk}{Spin glasses and other random magnets}
\pacs{64.60.Ht}{Dynamic critical phenomena}
\begin{document}
\maketitle
\begin{abstract}
The non-equilibrium ageing behaviour of the three-dimensional and 
four-dimensional critical Ising
spin glass is studied for both binary and gaussian disorder. 
The same phenomenology of the time-dependent scaling as in non-disordered 
magnets is found but the non-equilibrium exponents and the universal limit 
fluctuation-dissipation ratio depend on the distribution of the 
coupling constants. 
\end{abstract}

Understanding the kinetics of spin systems or glasses 
after a rapid quench from an initial disordered state 
continues to pose challenging problems. A key insight has been the 
observation that many of the apparently erratic and history-dependent 
properties of such systems can be organized in terms of
a simple scaling picture \cite{Stru78}. 
The simplest way to interpret this is by assuming the existence of a single 
time-dependent length-scale $L(t)$ in the problem.

For simple magnets without disorder, one typically finds a power-law
scaling $L(t)\sim t^{1/z}$, where $z$ is the dynamical exponent. 
It has been realized that the ageing behaviour is more fully
revealed in observables such as the two-time autocorrelation function 
$C(t,s) := \langle \phi(t) \phi(s) \rangle$ 
or the two-time linear autoresponse function  
$R(t,s) := \left.{\delta \langle \phi(t)\rangle}/{\delta h(s)}\right|_{h=0}$,
where $\phi(t)$ denotes the time-dependent order-parameter, $h(s)$ is the
time-dependent conjugate magnetic field, $t$ is referred to as {\em observation 
time} and $s$ as {\em waiting time}. One says that the system undergoes
{\em ageing} if $C$ or $R$ depend on both $t$ and $s$ and not merely on the
difference $\tau=t-s$. For simple magnets, these two-time functions are 
expected to show dynamical scaling in the ageing regime 
$t,s\gg t_{\rm micro}$ and $t-s\gg t_{\rm micro}$,
where $t_{\rm micro}$ is some microscopic time scale. Then 
\BEQ \label{gl:SkalCR}
C(t,s) = s^{-b} f_C(t/s) \;\; , \;\; R(t,s) = s^{-1-a} f_R(t/s)
\EEQ
where $a,b$ are non-equilibrium exponents. 
The scaling functions $f_{C,R}(y)$ should satisfy the following asymptotic
behaviour
\BEQ \label{gl:lambdaCR}
f_C(y) \sim y^{-\lambda_C/z} \;\; , \;\; 
f_R(y) \sim y^{-\lambda_R/z}
\EEQ
as $y\to\infty$ and where $\lambda_C$ and $\lambda_R$, respectively, are known
as the autocorrelation \cite{Fish88,Huse89} and autoresponse exponents
\cite{Pico02}. In what follows, we shall concentrate on quenches {\em onto} the
critical point. Then for simple ferromagnets 
$a=b=2\beta/\nu z=(d-2+\eta)/z$ where $\beta,\eta,\nu$ 
are standard equilibrium critical exponents. Furthermore, for spatially 
short-ranged initial correlations, one has $\lambda_C=\lambda_R$. These 
expectations on critical ageing have been confirmed in a large variety of 
models through simulations, exact solution or field-theory calculations,
see \cite{Cugl02,Godr02,Henk04,Cala04} for recent reviews. 
Critical ageing (\ref{gl:SkalCR}) also 
occurs in systems without detailed balance such as the contact process, but
now with the exponent relations $1+a=b=2\beta/\nu_{\perp}z$ while 
the autocorrelation and autoresponse exponents still co\"{\i}ncide 
\cite{Rama04,Enss04}.

It is natural to try and see whether these results might be extended to glassy 
systems. In this letter, we shall concentrate on the Ising spin glass, with a 
static Hamiltonian ${\cal H} = - \sum_{(i,j)} J_{i,j} \sigma_i \sigma_j$. 
Here $\sigma_i=\pm 1$ are the usual Ising spins and the nearest-neighbour
couplings $J_{i,j}$ are random variables. We shall consider (i) a binary
distribution where $J_{i,j} = \pm 1$ and (ii) a gaussian distribution of the
$J_{i,j}$ with zero mean and variance one.
The dynamics of the model is given by a master equation where the
rates are chosen according to heat-bath dynamics. It is 
now established \cite{Kawa04} that this model undergoes in $d>2$ dimensions an
equilibrium phase-transition between a paramagnetic and a frustrated 
spin-glass phase. On the other hand, when considering a quench below
the spin-glass critical temperature $T_c$, there has been considerable
debate on the precise relationship between the relevant time and length
scales. It has been attempted to summarize the present state of knowledge
into the form \cite{Bouc01}
\BEQ \label{gl:tL}
t(L) \sim L^z \exp\left(
\frac{\Delta_0}{T} \left(\frac{L}{\xi(T)}\right)^{\psi}
\right)
\EEQ
where $\Delta_0$ is an energy scale of order $T_c$, $\psi$ is a barrier
exponent and $\xi(T)$ is the equilibrium correlation length at 
temperature $T$. This form has been
used to fit successfuly simulational data in the $3D$ and $4D$ Edwards-Anderson 
model \cite{Bouc01,Bert02}. 
Although the typical length scales are merely of the 
order of a few lattice sizes, see e.g. \cite{Yosh02}, the relaxation times are
sufficiently large for a dynamical scaling to set in. We shall return 
to this below.  While this expression, if correct, 
points towards a cross-over behaviour between a simple power-law
scaling and an exponential scaling as would follow from the droplet model, 
it also suggests that at criticality, simple power-law scaling should prevail.
The determination of the exponent $z$ in spin glasses and 
in real materials is a much-studied topic,  
see \cite{Kawa04} and references therein.
It is the aim of this letter to further test this idea in the critical 
Ising spin-glass, paying special attention to the dynamical scaling 
behaviour of two-time functions in the ageing regime.
A similar analysis was recently performed in a damage-spreading context
\cite{Ozek01}. 
{}From now on we consider the three- and the four-dimensional
Ising spin glass quenched to $T=T_c$ (the assumed values of $T_c$ are 
listed in table~\ref{tab1}) from a fully disordered initial state. 
We shall study the scaling
behaviour of the magnetic autocorrelation 
$C(t,s) = {\cal N}^{-1}\sum_i \left\langle\sigma_i(t)\sigma_i(s)\right\rangle$
and of its associated linear response with respect to an external magnetic
field $h$. As usual, since response functions are too noisy to be measured
directly, we study instead the thermoremanent magnetization by turning
on the magnetic field immediately after the quench at time $t=0$ and keeping
it until after the waiting time $s$ has elapsed. 
The magnetization measured at a later time $t$ is
related to the linear response function by
$M(t,s) = h \int_{0}^{s} \!\D u\, R(t,u)$.
The systems simulated contained $50^3$ and $20^4$ spins, respectively. 
Some other system sizes were also briefly considered in order to check 
against finite-size effects. The autocorrelation data discussed in the 
following have been obtained after averaging over typically a few 
thousand different bond distributions. For the thermoremanent magnetization 
we averaged over at least ten thousand samples with different realizations 
of the couplings.

%%----------------------------------------------------------------------------%%
\begin{figure}[t]
\centerline{\epsfxsize=3.9in\epsfbox
{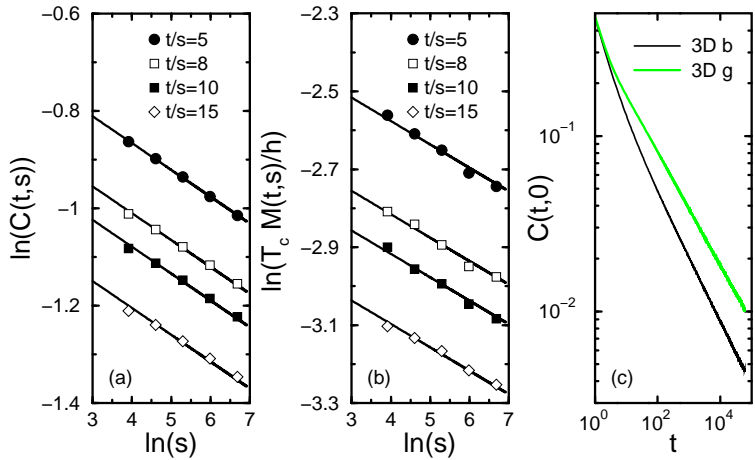}
}
\caption{Scaling of (a) the autocorrelation function $C(t,s)$ and
(b) the thermoremanent magnetization $M(t,s)$ of the $3D$ binary 
Edwards-Anderson spin glass at criticality. The full curves are
$C(t,s) = c_0 s^{-b}$ and $M(t,s) = m_0 s^{-a}$, where $c_0,m_0$ were
fitted to the numerical data. Here and in the following error bars are 
smaller than the symbols sizes. (c) Temporal evolution of the
autocorrelation $C(t,0)$ in the three-dimensional cases (b: binary,
g: gaussian). 
\label{Bild1}}
\end{figure}
%%----------------------------------------------------------------------------%%

We first study the scaling as a function of the waiting time $s$,
with $t/s$ fixed. In figure~\ref{Bild1} we show data for the $3D$ 
binary case. We observe that both correlation (figure~\ref{Bild1}a)
and integrated response (figure~\ref{Bild1}b) are
consistent with a power-law scaling. Similar results were obtained
also in $4D$ and for the gaussian case. In table~\ref{tab1} our 
results for the exponents $a$ and $b$ are listed. Their values are close to
results inside the spin-glass phase but near to $T_c$ \cite{Kisk96}. 

We observe that $a=b$ within our numerical errors for a fixed
choice of the distribution of the couplings. However, the results for  
binary and gaussian distributions are different.

In order to check that the results of table~\ref{tab1} relate to the 
true long-time scaling regime, we compare with the expected relation
$a=b=\beta_{\rm EA}/(\nu z)=(d-2+\eta_{\rm EA})/(2z)$ where 
$\beta_{\rm EA},\eta_{\rm EA}$ are the equilibrium critical exponents of the
Edwards-Anderson order parameter which is quadratic in the magnetization. 
Their values were measured many times. For the $3D$ binary case we quote 
$\eta_{\rm EA}=-0.225(25)$ and $z=5.65(15)$ \cite{Mari02} or
$\eta_{\rm EA}=-0.337(15)$ \cite{Ball00}, leading to $a=0.068(4)$ 
and $0.059(3)$, respectively. For the $3D$ gaussian case 
$\eta_{\rm EA}=-0.42(3)$ and
$z=6.45(10)$ \cite{Camp00,Katz04} leading to $a=0.044(3)$ whereas the result
$\eta_{\rm EA}=-0.36(6)$ \cite{Mari98} gives $a=0.049(5)$. 
In four dimensions one has for the binary case $\eta_{\rm EA}=-0.31(1)$
and $z=4.45(10)$ \cite{Bern97} leading to $a=0.19(1)$. Finally, 
for the $4D$ gaussian case $\eta_{\rm EA}=-0.35(5)$ \cite{Pari96} 
or $\eta_{\rm EA}=-0.44(2)$ \cite{Camp00,Katz04}
and $z=4.9(4)$ \cite{Camp00,Katz04} which gives $a=0.17(1)$ or $a=0.16(1)$. 
These results for $a$ agree very well with the values
extracted from our dynamical simulations in three dimensions and 
for the $4D$ binary case.
For the $4D$ gaussian case our value for $a$ is substantially larger 
than the value obtained
when combining the literature values of $\eta_{\rm EA}$ and $z$. 
The origin of this discrepancy in not clear to us.

A second non-equilibrium critical exponent is given by the power-law 
decay of the autocorrelation
for long times. As shown in figure~\ref{Bild1}c the 
quantity $C(t,0)$ displays a power-law
behaviour over more than three decades, making a very precise determination 
of the exponent $\lambda_C/z$ possible, see \cite{Huse89}, but which
does depend on the distribution of the $J_{ij}$. 
We have checked that the distribution-dependence of
$\lambda_C$ does not simply result from inaccurate determinations of $T_c$. 
In $3D$, using $T_c=1.14$ \cite{Ball00} we find $\lambda_C/z=0.348(5)$ for
the binary case and using $T_c=0.95$ \cite{Mari98}, we find 
$\lambda_C/z=0.335(5)$ for the gaussian case. These values still appear to be 
significantly different from each other. 

Summarizing, we have found
evidence that the dynamic universality class may depend on the choice
of the distribution of the coupling constants. For the equilibrium transition,
even the very existence of a phase-transition in $3D$ was questioned for a
long time, see \cite{Kawa04} and the question of its universality remains
controversial. For example, numerical and experimental evidence against
universality was presented in \cite{Camp00,Katz04} while a 
recent high-temperature
study of the Edwards-Anderson susceptibility with four symmetric random
distributions asserted the universality of the exponent $\gamma$ \cite{Dabo04}
in four to eight dimensions.

%%++++++++++++++++++++++++++++++++++++++++++++++++++++++++++++++++++++++++++++++
\begin{table}[t]
\caption{Critical temperature and universal 
nonequilibrium quantities of the critical 
Ising spin glass, for both binary and gaussian distributions of the 
nearest-neighbour couplings. In four dimensions even our longest runs 
did not permit us to reliably determine $\lambda_R/z$.
\label{tab1}}
\begin{tabular}{|c|lc|ccccc|}  \hline
    & \multicolumn{7}{|c|}{binary}                    \\ \hline
$d$ & \multicolumn{2}{c|}{$T_c$} & $a$ & $b$ & $\lambda_C/z$ & $\lambda_R/z$ & $X_\infty$ \\ \hline
3   & 1.19 & \cite{Mari02} & 0.060(4) & 0.056(3) & 0.362(5)~ & 0.38(2) & 0.12(1) \\
4   & 2.0  & \cite{Huku99} & 0.18(1)~ & 0.17(1)~ & 0.615(10) &     -   & 0.19(1) \\ \hline
 & \multicolumn{7}{|c|}{gaussian} \\ \hline
$d$ & \multicolumn{2}{c|}{$T_c$} & $a$ & $b$ & $\lambda_C/z$ & $\lambda_R/z$ & $X_\infty$ \\ \hline
3   & 0.92 & \cite{Camp00} & 0.044(1) & 0.043(1) & 0.320(5) & 0.33(2) & 0.09(1) \\
4   & 1.78 & \cite{Camp00} & 0.22(1)~ & 0.23(1)~ & 0.68(1)~ & -       & 0.16(1) \\ \hline
\end{tabular}
\end{table}
%%++++++++++++++++++++++++++++++++++++++++++++++++++++++++++++++++++++++++++++++

%%----------------------------------------------------------------------------%%
\begin{figure}[tb]
\centerline{\epsfxsize=5.6in\epsfbox
{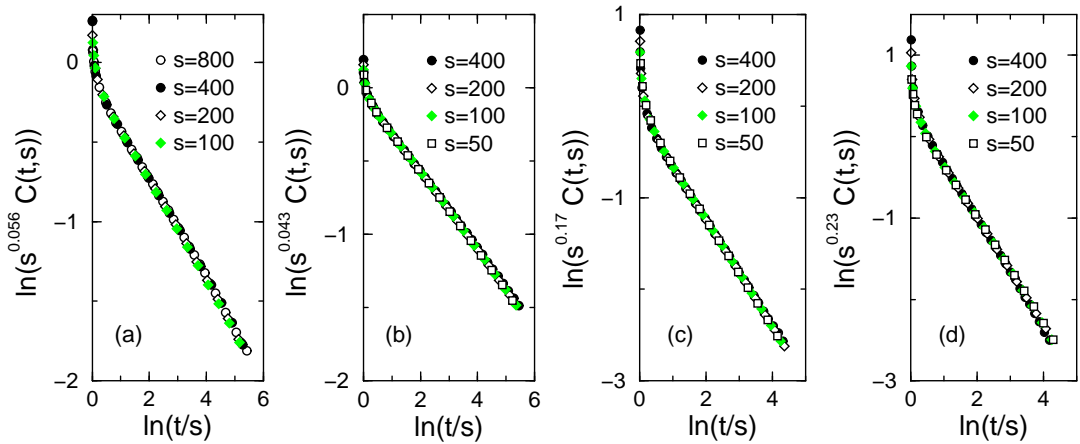}
}
\caption{Scaling of the autocorrelation in the critical 
Edwards-Anderson spin-glass with the following dimensions and
couplings: (a) $3D$ binary, (b) $3D$ gaussian, (c) $4D$ binary and (d)
$4D$ gaussian. 
\label{Bild2}}
\end{figure}
%%----------------------------------------------------------------------------%%

Next, we turn to the form of the scaling functions themselves. 
In figure~\ref{Bild2}, we show the two-time scaling of the spin-spin
autocorrelator, for several waiting times $s$. 
It is clear that $C(t,s)$ does not merely depend on the 
time difference $t-s$ and therefore the system ages. We find in all cases
a nice collapse of the rescaled autocorrelator $s^b C(t,s)$ compatible with 
a simple power-law scaling $L(t)\sim t^{1/z}$. 

%%----------------------------------------------------------------------------%%
\begin{figure}[tb]
\centerline{\epsfxsize=3.9in\epsfbox
{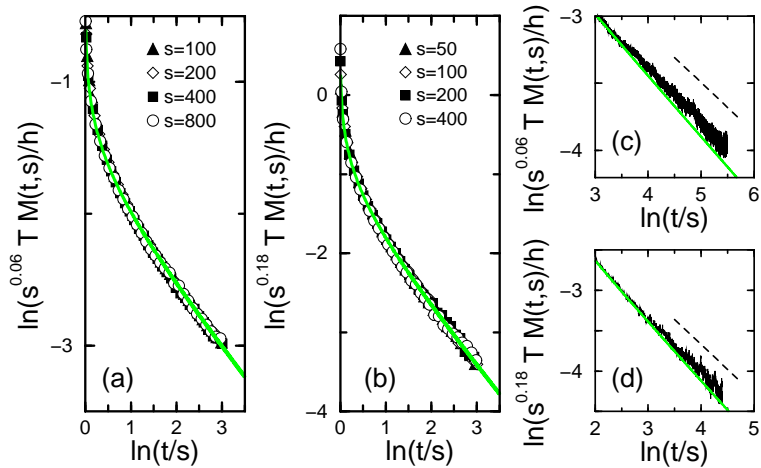}
}
\caption{Scaling of the thermoremanent magnetization in the
Edwards-Anderson spin glass at criticality with binary couplings in (a) $3D$ 
and (b) $4D$. The full curve is the prediction
(\ref{gl:fM}) of local scale-invariance, see text. 
The results of very longs runs for a single waiting time
are shown in (c) $3D$ with $s=100$ and (d) $4D$ with $s=25$. 
\label{Bild3}} 
\end{figure}
%%----------------------------------------------------------------------------%%

In a similar way, in figure~\ref{Bild3} we show the two-time scaling of the
integrated response. Again, we observe a very nice collapse of the data
in terms of a simple power-law scaling. The fact that both the autocorrelator
and the thermoremanent magnetization can be described in terms of such a
power-law scaling is evidence in favour of the time-dependent length-scale
(\ref{gl:tL}). Closer inspection of figure~\ref{Bild3}cd, however, reveals
an unexpected subtlety. In principle, one would like to extract an 
exponent $\lambda'_R/z$ from the slopes in that figure. 
It turns out that the values of $\lambda'_R/z$ thus obtained 
are significantly different (3D binary: 0.45, 3D gaussian: 0.41,
4D binary: 0.72, 4D gaussian: 0.76) from the
ones found before for $\lambda_C/z$. Indeed, if one goes to larger values
of $y=t/s$ as is shown in figure~\ref{Bild3}cd, we find that
our data are systematically above the asymptotic power-law 
$\sim (t/s)^{-\lambda_R'/z}$ which we obtained from smaller values of $t/s$.
This passage from an effective exponent $\lambda_R'/z$ 
at intermediate values of the scaling variable $y$ to the truly
asymptotic value $\lambda_R/z$ at larger values of $y$ has also been 
observed in the critical ageing of the ferromagnetic Ising model with
Kawasaki dynamcis \cite{Godr04,Sire04}. We shall return to a more quantitative 
discussion of $M(t,s)$ below.

%%----------------------------------------------------------------------------%%
\begin{figure}[t]
\centerline{\epsfxsize=3.9in\epsfbox
{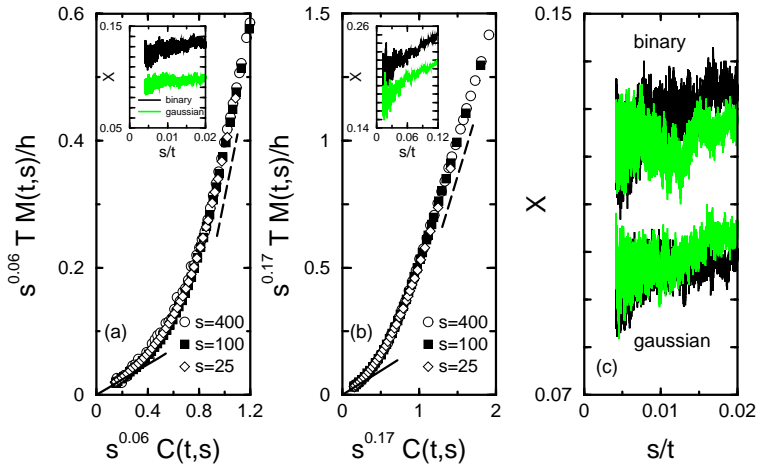}
}
\caption{Determination of the fluctuation-dissipation ratio in the 
critical Edwards-Anderson spin glass with binary couplings in (a) $3D$ and
(b) $4D$. The dashed lines have slope one and show the quasi-equilibrium
behaviour while the slopes of the full lines give the limit 
fluctuation-dissipation ratio $X_{\infty}$. The insets show the function
$X :=T_c M(t,s)/(h C(t,s))$ as a function of $s/t$ for both the binary
and the gaussian cases.  In (c) we show $X$ in $3D$ for the choices of
the distribution and of $T_c$: 
binary with $T_c=1.19$ (black) and $T_c=1.14$ (gray), 
gaussian with $T_c=0.92$ (black) and $T_c=0.95$ (gray).
\label{Bild4}}
\end{figure}
%%----------------------------------------------------------------------------%%

In simple ferromagnets, ageing occurs out of equilibrium. The distance from 
the equilibrium state may be measured through the fluctuation-dissipation
ratio 
\BEQ
X(t,s) = \frac{T R(t,s)}{\partial C(t,s)/\partial s} = \hat{X}(t/s)
\EEQ
In particular, the limit fluctuation-dissipation ratio 
$X_{\infty}=\lim_{x\to\infty} \hat{X}(x)$ is expected to be an {\em universal} 
number \cite{Godr02}.  
In figure~\ref{Bild4} we display the fluctuation-dissipation relation between
the scaled autocorrelation and thermoremanent magnetization in both $3D$ and
$4D$. For small time differences $t-s$, 
the autocorrelation $C(t,s)$ is large and we are in a
quasiequilibrium regime with a fluctuation-dissipation ratio $X(t,s)\approx 1$. 
On the other hand, for well-separated times $s$ and $t$, the systems moves out
of equilibrium and we can read off the limit fluctuation-dissipation ratio
$X_{\infty}$. We remark that since we had to use the {\em scaled} forms
$s^a C(t,s)$ and $s^a M(t,s)$ rather than the unscaled observables, $X(t,s)$
cannot merely depend on the value of the autocorrelation, in distinction with
what occurs in the mean-field theory of spin glasses. 

Finally, in figure~\ref{Bild4}c we display the ratio $T_c M(t,s)/(h C(t,s))$ 
in $3D$ as a function of $s/t$ and for several choices of $T_c$. 
For widely separated times, $s/t\to 0$ and this
ratio should converge to the limit fluctuation-dissipation ratio $X_{\infty}$. 
We therefore find (see table~\ref{tab1}) that 
$X_{\infty}$ apparently depends on the distribution of
the couplings as well, and independently of the assumed $T_c$. 

The qualitative dynamical scaling behaviour of the 
magnetic two-time observables is quite analogous to the one of
the conceptually simpler magnets, up to modified values of the
critical nonequilibrium exponents. We now inquire into the form of the 
scaling function of the thermoremanent magnetization. 
It is known that by extending dynamical 
scaling eq.~(\ref{gl:SkalCR}) to a {\em local} scale-invariance with 
infinitesimal local scale-transformations 
$t\mapsto (1+\eps)^z t$, $\vec{r} \mapsto (1+\eps)\vec{r}$
(with an infinitesimal $\eps=\eps(t,\vec{r})$ which  
depends on both time and space), the form of $f_R$ can be found from the
requirement of covariance of $R(t,s)$ under these transformations, 
leading to \cite{Henk02,Henk01}
$f_R(y) \sim  y^{1+a-\lambda_R/z} \left( y-1 \right)^{-1-a}$. 
This prediction has been reproduced 
in many spin systems quenched to a temperature $T\leq T_c$ and whose
dynamics is described by a master 
equation \cite{Henk02,Henk01,Henk03b,Henk04,Pico04,Plei04,Enss04}. 
Hence, we expect $M(t,s) = M_0 s^{-a} f_M(t/s)$, where 
\BEQ \label{gl:fM}
f_M(y) = y^{-\lambda_R/z} {_2F_1}\left(1+a,-a+{\lambda_R}/{z};
1-a+{\lambda_R}/{z};{1}/{y}\right) 
\EEQ
with a normalization constant $M_0$. 
At $T=T_c$, however, a second-order $\eps$-expansion produces a deviation from 
that prediction \cite{Cala04}.
As discussed previously we observe in critical spin glasses a passage from
an effective exponents $\lambda'_R/z$ at intermediate values of $y=t/s$ to the
truly asymptotic value for larger $y$. Hence we cannot expect 
equation (\ref{gl:fM}) to describe the scaling function $f_M(y)$ 
for all values of $y=t/s$.
In figure~\ref{Bild3}ab we compare the numerical data, 
in both $3D$ and $4D$ with the prediction (\ref{gl:fM}) where we have
inserted the values of the exponents $a$ and $\lambda'_R/z$
which have been determined earlier. 
While we find a nice agreement of the prediction (\ref{gl:fM}) of local
scale-invariance with our data for $y=t/s$ not too large, for very large
arguments the precise behaviour of the scaling function $f_M(y)$ cannot
be fully reproduced.   

In summary, we have studied the dynamical scaling behaviour of 
critical three- and four-dimensional Ising spin glasses in the ageing 
regime. We find for these disordered systems evidence of a dynamical power-law
scaling analogous to non-disordered critical magnets.
The measured values of several independent universal non-equilibrium 
critical quantities suggest that the dynamic universality class may depend 
on the choice of the distribution of the random coupling constants.

\acknowledgments
We thank A.J. Bray, C. Godr\`eche, J.-M. Luck, M. Moore and 
F. Ricci-Tersenghi for discussions. 
This work was supported by the Bayerisch-Franz\"osisches Hochschulzentrum
(BFHZ), by CINES Montpellier (projet pmn2095), and by NIC J\"{u}lich
(Projekt Her10). MP acknowledges the support by 
the Deutsche Forschungsgemeinschaft through grant no. PL 323/2.

%%++++++++++++++++++++++++++++++++++++++++++++++++++++++++++++++++++++++++++++++

%%%%%%%%%%%%%%%%%%%%%%%%%%%%%%%%%%%%%%%%%%%%%%%%%%%%%%%%%%%%%%%%%%%%%%%%%%%%%%%%

\end{document}